\documentclass[acus]{JAC2003}

\usepackage{graphicx}
\usepackage{booktabs}

\begin{document}
\title{High flux polarized gamma rays production: first measurements with a four-mirror cavity at the ATF}

\author{N. Delerue\thanks{delerue@lal.in2p3.fr}, J. Bonis, I. Chaikovska, R. Chiche, R. Cizeron, M. Cohen,\\ J. Colin, P. Cornebise, D. Jehanno, F. Labaye, M. Lacroix, R. Marie,\\ Y. Peinaud, V. Soskov, A. Variola, F. Zomer,\\{\em LAL, Universit\'e Paris-Sud XI, Orsay, France}\\
E. Cormier, {\em CELIA, Universit\'e de Bordeaux, Bordeaux, France}\\
R. Flaminio, L. Pinard, {\em LMA, Lyon, France}\\
S. Araki,  S. Funahashi, Y. Honda, T. Omori, H. Shimizu, T. Terunuma, J. Urakawa, {\em KEK, Tsukuba, Japan}\\
T. Akagi, S. Miyoshi, S. Nagata, T. Takahashi, {\em Hiroshima University, Hiroshima, Japan}}

\maketitle

\begin{abstract}
The next generation of e+/e- colliders will require a very intense flux of gamma rays 
to allow high current polarized positrons to be produced. This can be achieved by converting polarized high energy photons in polarized pairs into a target.
In that context, an optical system consisting of a laser and a four-mirror passive Fabry-Perot cavity has recently been installed at the Accelerator Test Facility (ATF) at KEK to produce a high flux of polarized gamma rays by inverse Compton scattering. In this contribution, we describe the experimental system and present preliminary results. An ultra-stable four-mirror non planar geometry has been implemented to ensure the polarization of the gamma rays produced. A fiber amplifier is used to inject about 10W in the high finesse cavity with a gain of 1000. A digital feedback system is used to keep the cavity at the length required for the optimal power enhancement. Preliminary measurements show that a flux of about $4\times10^6$~$\gamma$/s with an average energy of about 24 MeV was generated. Several upgrades currently in progress are also described.
\end{abstract}

\section{Introduction}

Future electron-positrons colliders will require a flux of positrons much higher than what can be produced with current methods. One of the scheme that has been proposed to achieve such flux is to use an intense flux of polarized high energy gamma rays~\cite{omori} which can be converted into polarized electron-positron pairs. Although promising this method requires the demonstration that such high intensity flux of gamma rays can be produced. Polarized positrons can be produced by Compton interaction between an electron beam and a laser. However given the low cross-section of this process, it is desirable to recycle both the photons and the electrons by, for instance, using an electron ring and a Fabry-Perot cavity (FPC).
We decided to investigate such a scheme by installing a four-mirror FPC on the Accelerator Test Facility (ATF)~\cite{ATF}  damping ring at KEK in Japan.

\section{Experimental setup}

\subsection{Laser system}

\begin{figure}[hbt]
    \centering
    \includegraphics*[width=85mm]{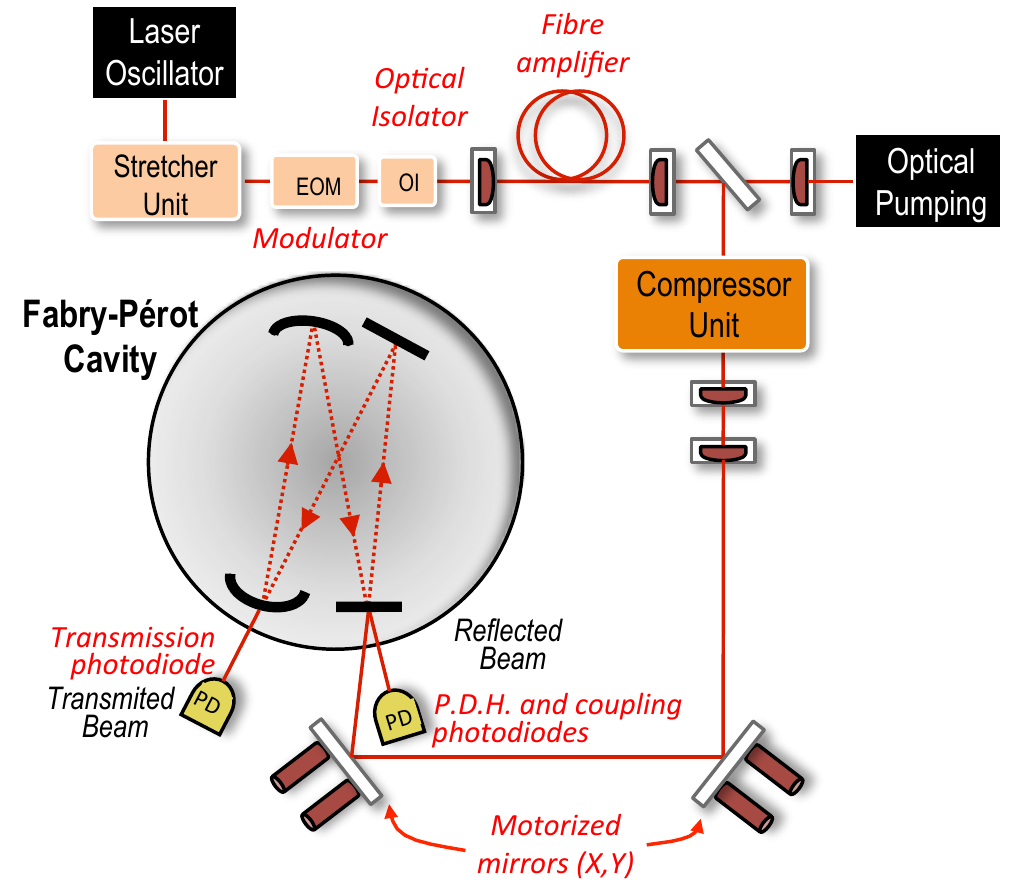}
    \caption{Optical layout of the experiment.}
\label{fig:optical_layout}
\end{figure}

The laser seed includes a modified commercial oscillator emmiting at 1031 nm with a repetition rate of 178.5 MHz. We use a  standard chirp pulse amplification (CPA) architecture in order to limit the non-linearities in the amplifier. The amplifier is built around a commercial microstructure Yb-doped double clad fiber. The system delivers recompressed pulses of around 60 ps and up to 50 W average power. The complete optical layout is shown on figure~\ref{fig:optical_layout}.

\subsection{Fabry-Perot cavity (FPC)}

\begin{figure}[htb]
   \centering
   \includegraphics*[width=85mm]{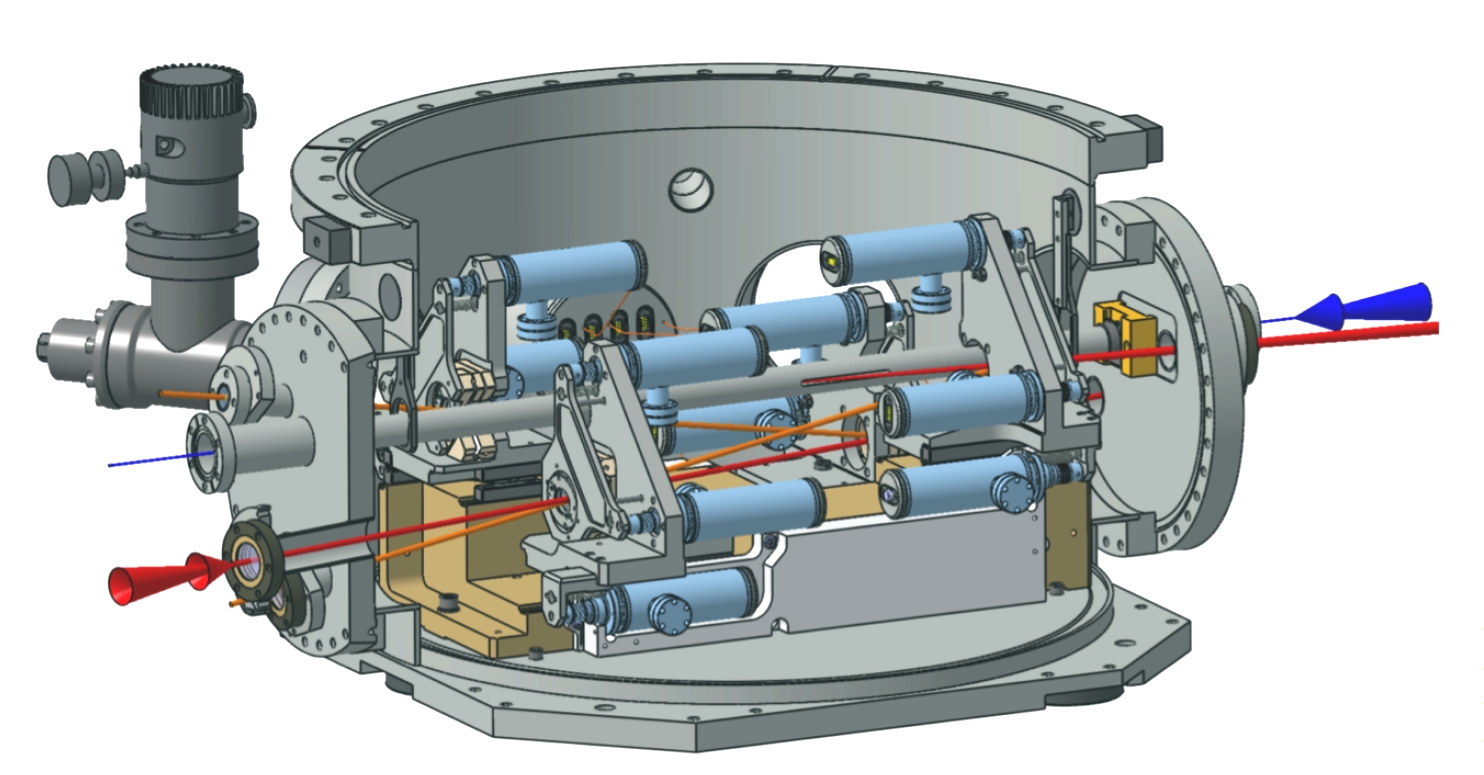}\vspace*{-3mm}
   \caption{Layout of the 4-mirror cavity. The electron beam is represented by the blue arrow and the blue line entering to the right of the cavity. The laser beam is represented by the red arrow and the red line entering to the left of the cavity. The blue cylinders near the mirrors contain the 12 actuators used to remotely adjust the X, Y and Z positions of the mirrors.  }
   \label{fig:cavity_geo}
\end{figure}

The FPC is built using the experience gained at LAL in previous experiments~\cite{DESY_paper,PLIC}.
The FPC is formed by 2 concave mirrors with a radius of curvature of 0.5m and 2 flat mirrors arranged in a non-planar tetrahedron geometry (see figure~\ref{fig:cavity_geo}). 
The two concave mirrors form a $52\mu m \times 76\mu m$ waist at which the laser beam crosses the electron beam with a collision angle of 8 degrees.
The mirrors have a very high reflectivity (1~-~1060~ppm for one of them and 1~-~330~ppm for the others) leading to a cavity finesse of the order of 3000 (that is a power enhancement of about 1000).
 The duration of a round trip in the FPC is 5.6~ns. Great care had to be taken while designing the cavity to ensure that it is compatible with the ultra-high vacuum requirements of the ATF ($3\times 10^{-8}$ mbar).

\begin{figure}[htb]
   \centering
   \hspace*{-1cm}\includegraphics*[width=90mm]{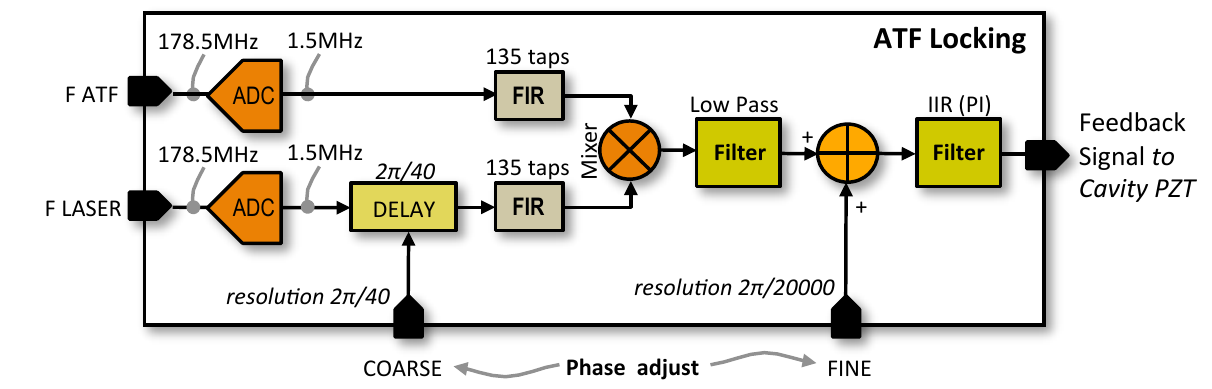}\vspace*{-3mm}
   \caption{Digital phase lock loop used to synchronize the FPC with the ATF clock.}
   \label{fig:digital_PLL}
\end{figure}

All the mirrors are mounted on actuators carts or tables that allow a fine adjustment of the cavity length. Furthermore one of the mirrors is mounted on a piezoelectric transducer (PZT) that allows to synchronize the length of the FPC with the repetition frequency of the ATF by using a digital phase lock loop as shown on figure~\ref{fig:digital_PLL}.

\begin{figure}[htb]
   \centering
   \includegraphics*[width=70mm]{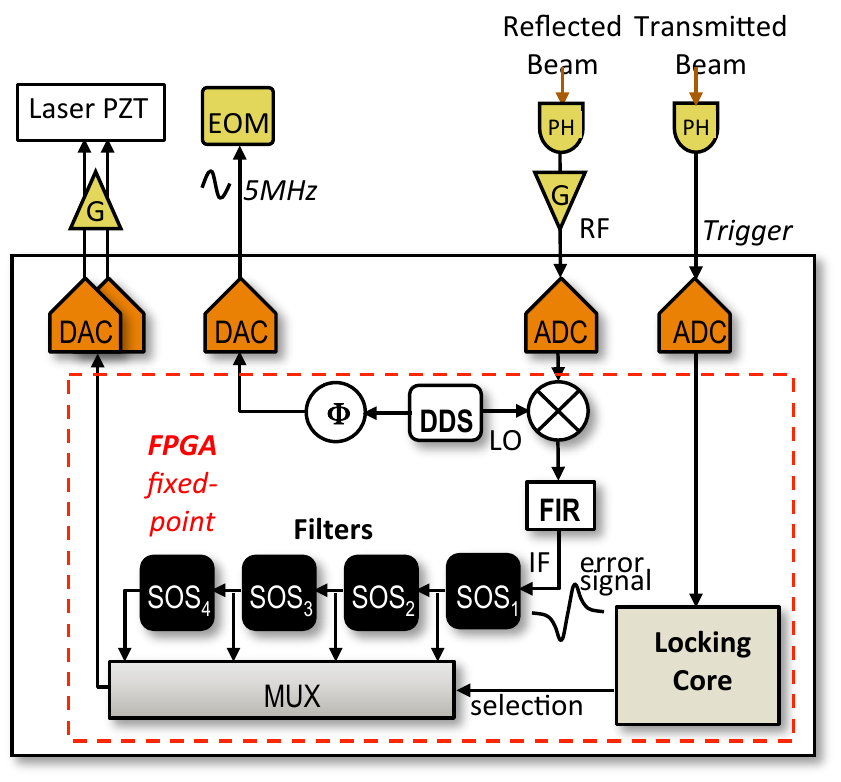}\vspace*{-3mm}
   \caption{Digital feedback to adjust the laser of the laser cavity depending on the PDH signal.}
   \label{fig:digital_PDH}
\end{figure}

In order to constructively stack the pulses in the FPC, it is mandatory to control the pulse to pulse phasing with an interferometric accuracy. 
This is done using the Pound-Drever-Hall method~\cite{PDH} implemented in a VIRTEX-II FPGA board as shown on figure~\ref{fig:digital_PDH}. The error signal generated by this method is used to adjust the length of the laser oscillator.

Using this double digital feedback system we have been able to keep the FPC locked on the ATF and the laser oscillator locked on the FPC for several hours.

\subsection{ATF}

A detailed description of the ATF at KEK can be found in the literature~\cite{ATF}. At the ATF the 1.28~GeV electron bunches  are separated by 2.8~ns. Although various fill patterns can be achieved in the ATF damping ring (DR), the data presented here were taken in single bunch single train mode in which the same electron bunch returns to the same position after 165 RF buckets (one complete revolution of the ATF DR equals to 462~ns). Given that a round trip in the FPC is 5.6~ns this means that electrons collide with the laser pulse stored in the FPC every other turn.


\subsection{Calorimeter}

During the collisions the gamma rays  are detected using a fast scintillation detector made of  barium fluoride ($BaF_2$) coupled with a Photomultiplier Tube (PMT). To eliminate the slow component of scintillation an optical filter is installed in front of the PMT. The data acquisition is performed using a LeCroy WS454 oscilloscope (1GS/s, 500 MHz bandwidth).  Geant4 simulations have been used to study this calorimeter.


\section{Data taking and data analysis}

The FPC was commissioned in October 2010 and Compton collisions were recorded on the first attempt. We present here a brief analysis of some of the data taken.

During data taking we record the waveforms from the PMT as well as the 357 MHz ATF clock and laser power transmitted by FPC measured by a photodiode. A full waveform contains approximately 200 000 samples spaced by  1 ns. On the day where the data presented here were taken the average power stored in the cavity was 160~W.

The 357 MHz clock is used to define the beginning of the periods in time during which the collisions occurred. So, the length of the period is naturally 924 ns. For each period we define the gate which contains the Compton peak  to calculate consequently its height and  integral. Different quality cuts such as a cut on the shape of the electronic pulse, a cut on the arrival time of the signal etc. were applied to restrict the analysis to a high purity sample.

\begin{figure}[htb]
   \centering
   \includegraphics*[width=62mm]{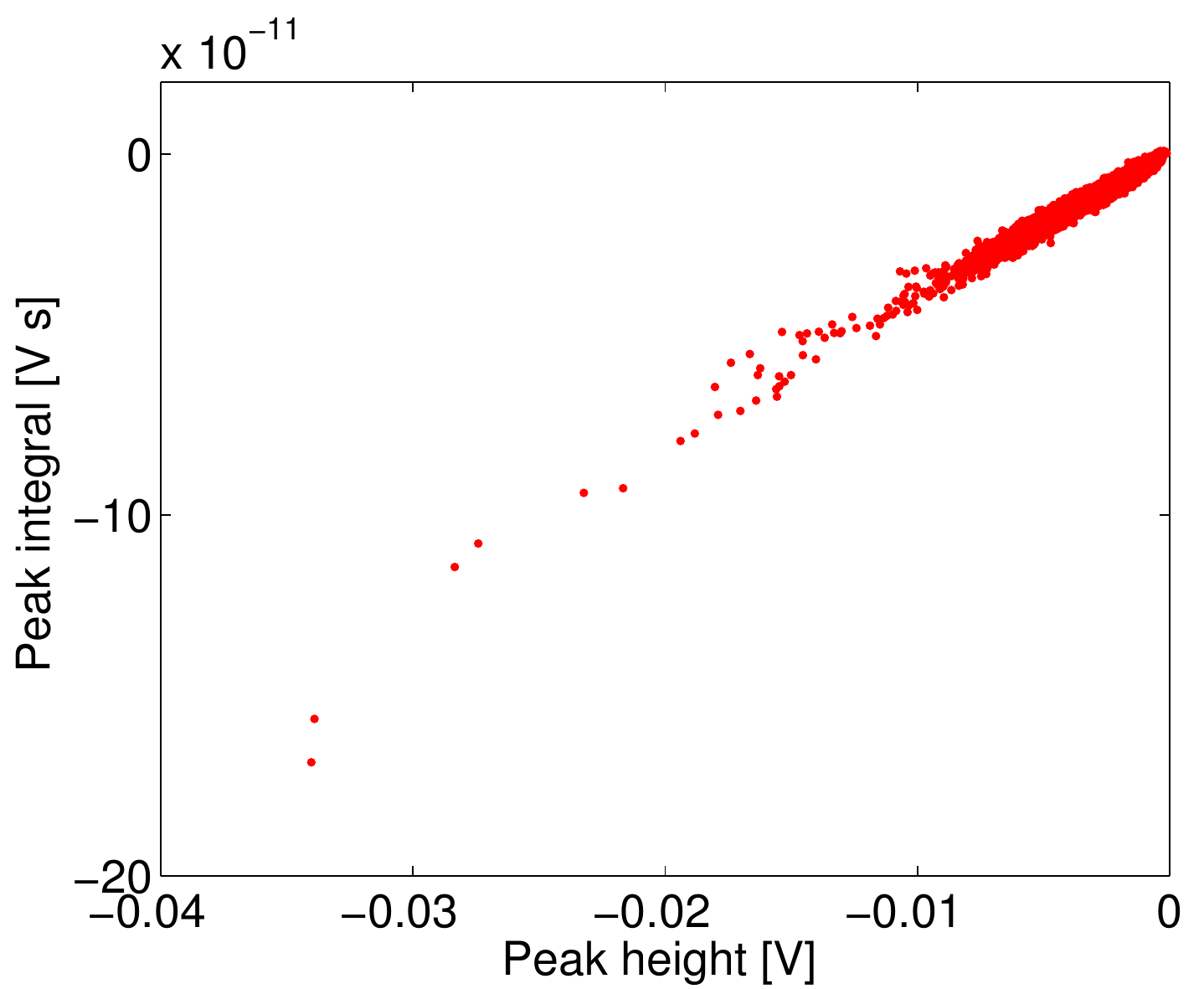}\vspace*{-3mm}
   \caption{Peak height vs. peak integral for the high purity data sample.}
   \label{fig:slope}
\end{figure}

The shape of the detector's response creates a linear relation between the total charge and the maximum charge measured. This can be seen as a correlation between the calculated peak height and peak integral as shown on figure~\ref{fig:slope}.  

\vspace*{-4mm}

\section{Results}

\begin{figure}[htb]
   \centering
   \includegraphics*[width=62mm]{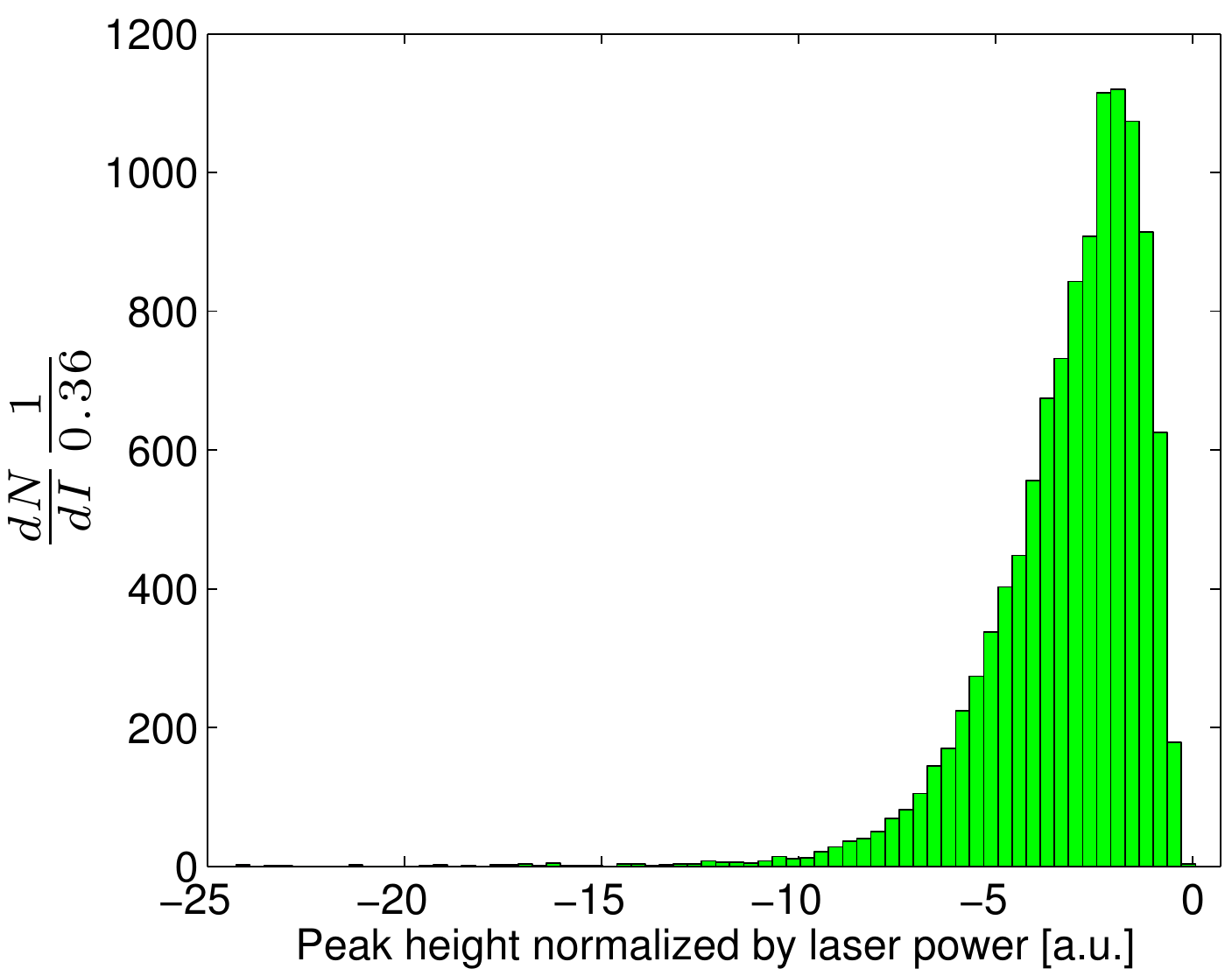}\vspace*{-3mm}
   \caption{Gamma ray spectrum.}
   \label{fig:spectrum}
\end{figure}

The spectrum of the gamma rays  is shown on figure~\ref{fig:spectrum}. 
It corresponds to the distribution of the energy deposited in the calorimeter expressed by the peak heights normalized by the laser power. 

Since we took data over a wide range of power stored in the FPC, we can study the gamma ray production for different ranges of that power. 
\begin{figure}[htb]
   \centering
   \includegraphics*[width=62mm]{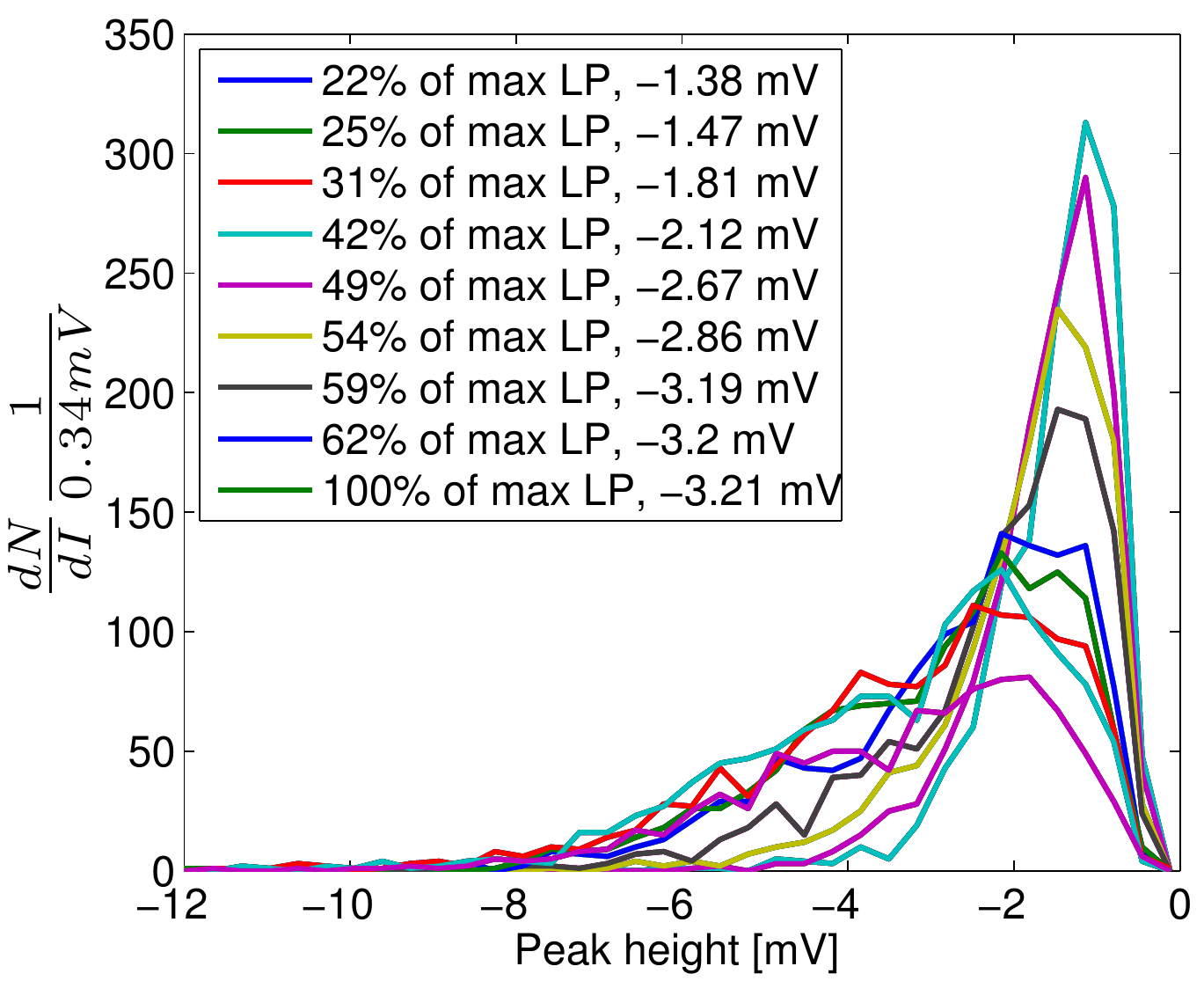}\vspace*{-3mm}
   \caption{Gamma spectrum for different FPC stored power. Different colors correspond to the spectra taken at different FPC powers. In the legend, the fraction of the maximum FPC stored power  and the mean of each of these spectra are shown.}
   \label{fig:laserbins}
\end{figure}
On figure~\ref{fig:laserbins}, the average of the peak height  distribution goes towards higher energy with increasing FPC power as expected.

Using cosmic rays for calibration, we can estimate the average number of scattered gammas per bunch crossing. According to this calibration, a peak height of 1 mV is equivalent to 34 MeV of energy deposited in the  calorimeter. Assuming the mean energy of scattered gammas to be 24 MeV, approximately 4 gammas are produced in average per bunch crossing (for an average laser power stored in the cavity of 160 W). As the repetition frequency of the collisions is about 1 MHz the flux of gamma rays achieved is about $4\times10^6$~$\gamma$/s (this does not take into account the 0.75 duty cycle of the ATF).

\section{Future plans}

Our experiment has been suspended due to the earthquake that struck Japan in March 2011 however we plan to resume soon. To increase the Compton flux we intend to replace the mirrors of the cavity by mirrors with a higher reflectivity, giving a finesse of about 30~000. We also plan to improve the laser system and to replace our gamma rays data acquisition chain with a faster one. 

Numerical simulations have shown that with the flux achieved so far we do not expect significant effects on the dynamics of the stored beam~\cite{Iryna}. However, with these future improvements, we intend to study the impact of Compton collisions on the beam dynamics in the DR.

This effort is an important step toward Compact X-ray sources such as ThomX~\cite{thomX} or polarized positrons source for the next generation of e+/e- colliders.

\end{document}